\documentclass[aps,pra,reprint,superscriptaddress,amsmath,amssymb]{revtex4-2}
\usepackage{graphicx}
\usepackage{xcolor}
\usepackage{float}
\usepackage{braket}

\usepackage{graphicx}
\usepackage{xcolor}
\usepackage[colorlinks=true,bookmarks=false,citecolor=black,urlcolor=black]{hyperref} 
\usepackage{float}
\usepackage{caption,subcaption}

\newcommand*{\rttensortwo}[1]{\bar{\bar{#1}}}

\begin{document}
\title{Enhanced vibrational optical activity by near-zero index chiral effective media}

\author{Ashis Paul} 
\email{ashiskumar.paul@univaq.it}
\affiliation{Department of Physical and Chemical Sciences, University of L'Aquila, Via Vetoio, 67100 L'Aquila, Italy}

\author{Matteo Venturi} 
\affiliation{Department of Physical and Chemical Sciences, University of L'Aquila, Via Vetoio, 67100 L'Aquila, Italy}

\author{Raju Adhikary}
\affiliation{Department of Physical and Chemical Sciences, University of L'Aquila, Via Vetoio, 67100 L'Aquila, Italy}

\author{Giovanna Salvitti}
\affiliation{Department of Physical and Chemical Sciences, University of L'Aquila, Via Vetoio, 67100 L'Aquila, Italy}

\author{Andrea Toma}
\affiliation{Istituto Italiano di Tecnologia, Via Morego 30, Genova 16163, Italy }

\author{Francesco Di Stasio}
\affiliation{Istituto Italiano di Tecnologia, Via Morego 30, Genova 16163, Italy }

\author{Hatice Altug}
\affiliation{Institute of Bioengineering, \'Ecole polytechnique f\'ed\'erale de Lausanne (EPFL), Lausanne 1015, Switzerland} 

\author{Paola Benassi}
\affiliation{Department of Physical and Chemical Sciences, University of L'Aquila, Via Vetoio, 67100 L'Aquila, Italy}

\author{Davide Tedeschi}
\affiliation{Department of Physical and Chemical Sciences, University of L'Aquila, Via Vetoio, 67100 L'Aquila, Italy}

\author{Carino Ferrante}
\affiliation{CNR-SPIN, c/o Dip.to di Scienze Fisiche e Chimiche, Via Vetoio,  L'Aquila 67100, Italy}

\author{Andrea Marini} 
\email{andrea.marini@univaq.it}
\affiliation{Department of Physical and Chemical Sciences, University of L'Aquila, Via Vetoio, 67100 L'Aquila, Italy}
\affiliation{CNR-SPIN, c/o Dip.to di Scienze Fisiche e Chimiche, Via Vetoio,  L'Aquila 67100, Italy}

\begin{abstract}
The enhancement of the inherently weak optical activity of solvated molecules by superchiral fields, crucial for detecting their chirality, is a research frontier of photonics and the basis of novel chiroptical detection schemes. Here, we show that an effective medium consisting of randomly dispersed metal-based nanoparticles embedded within an optically active solvated drug (aqueous reparixin) can enhance vibrational optical rotation and circular dichroism thanks to superchirality produced by slow light in near-zero index conditions. We evaluate from first principles the effective bianisotropic response of the bulk chiral effective medium, showing that, by adjusting the nanoparticles filling fraction, vibrational optical activity is greatly enhanced by a factor $\simeq 10^2-10^3$ at the near-zero index resonance. Our results are relevant for the development of innovative devices capable of detecting the chirality of low-volume samples, with applications in quantum chemistry and nanomedicine. 
\end{abstract}
\maketitle

\textit{Introduction}---Chirality, the handedness of a geometrical object with broken mirror symmetry, is crucial in diverse areas of science encompassing biology \cite{barron2008}, particle physics \cite{wu1957}, and optics \cite{mun2020}. Drug enantiomers--molecular forms with opposite handedness--present chirality-dependent interaction with biological tissues, affecting their functionality and toxicity \cite{nguyen2006}. Current chiral sensing techniques, e.g., nuclear magnetic resonance \cite{shundo2009} and high-performance chromatography \cite{okamoto2008}, enable enantiomer detection of ml drug volumes but are unsuitable for real-time analysis and on-chip integration schemes. Circular dichroism (CD) spectroscopy exploits the interaction of chiral molecules with circularly polarised light to detect the enantiomeric excess of solvated chiral mixtures \cite{ninabook}. However, for most chiral molecules, the dominant CD signal falls in the ultraviolet spectral range and is inherently weak \cite{tangprl}, preventing detection of low-volume samples. To overcome this challenge, diverse innovative ultrafast and nonlinear chiroptical spectroscopy techniques are currently being investigated to enhance chiroptical sensitivity, e.g., the exploitation of superchiral fields \cite{tangSC,ayusosmirnova2019SC,mohammadi2019SC,ayusosmirnova2022SC}, photoexcitation CD \cite{beaulieu2018PCD}, cavity-enhanced CD \cite{feisprl2020CAV},  and high harmonic generation \cite{meijerprlSHG1990,
smirnovaHHGPRX2019}. Moreover, nanophotonic platforms enable CD spectroscopy with enhanced chiroptical sensitivity, e.g., photonic nanocrystals capped with chiral molecules \cite{goborovfan2010NCmol,
goborovfan2012NCmol}, plasmonic nanostructures \cite{venturi2023,Raju2025}, chiral assemblies of plasmonic nanoparticles (NPs) \cite{fangoborov2010NPA,sloickgoborov2011NPA,Hentschel3DNPA2012,valev2013NS,mohammadi2018MS,jonesprlNPA2020}, dual nanodimers \cite{mohammadi2021NDM}, and metasurfaces \cite{solomon2018MS,baisiriMS2019}. These techniques rely on the chirality transfer from the molecule to the considered nanostructure, thanks to the creation of hotspots facilitated through dipole-dipole interactions \cite{goborov2011Cdtransfer,
zhangoborov2013CDtransfer,
nesterov2016NPA,
mohammadi2023CDtransfer}. In a complementary direction, chiral metamaterials have attracted a good deal of attention due to their strong chiral bi-anisotropic response enabling several applications such as broadband circular polarisation \cite{gansel2009PMM}, large optical activity (OA) \cite{gonokamiprlOA2005, KivsharGOAprb2014,
cuiGOA2014,GorkunovGOA2015}, asymmetric transmission \cite{MenzelASTprl2010}, and negative refraction \cite{zhangprl2009NR,plumprb2009NR,pendry2014NR}, to name a few. Near-zero-index (NZI) materials and  metamaterials, natural/artificial media exhibiting very small dielectric permittivity, enable to boost optical nonlinearities \cite{BoydAlam,ciattoni2016} thanks to the reduced group velocity in NZI conditions \cite{Ciattoni2013,Stockmann}. To the best of our knowledge, the exploitation of slow-light regime for enhancing chiroptical sensitivity remains hitherto unexplored.

In this Letter, we propose NZI chiral effective media (CEM) for enhancing the vibrational optical activity (VOA) of solvated chiral mixtures, embedding randomly dispersed metallic NPs, to attain superchirality thanks to slow-light regime in effective NZI conditions. In our calculations we focus on aqueous reparixin, a pharmaceutical molecule curing acute respiratory distress syndrome, including COVID-19 \cite{gorio2007,landonni2022}. We evaluate magneto-electric polarisabilities of metallic nanospheres (NSs) surrounded by aqueous reparixin in the quasi-static approximation and dilute molecular density conditions. Furthermore, we adopt the Maxwell-Garnett effective medium homogenisation technique to obtain the effective bianisotropic response of the bulk chiral CEM. We observe that the effective relative dielectric permittivity (RDP) of the CEM can be tuned by the medium filling fraction, enabling to achieve a vanishing RDP real part over a broad spectral range. Moreover, we find that, thanks to superchirality induced by NZI conditions, such CEMs exhibit enhanced vibrational CD (VCD) and optical rotatory dispersion (ORD). Our results shed light on the opportunities provided by nanophotonics for innovative chirality detection schemes of low-volume samples.
\begin{figure*}[t!]
	\centering
	\begin{center}
		\includegraphics[width=\textwidth ]{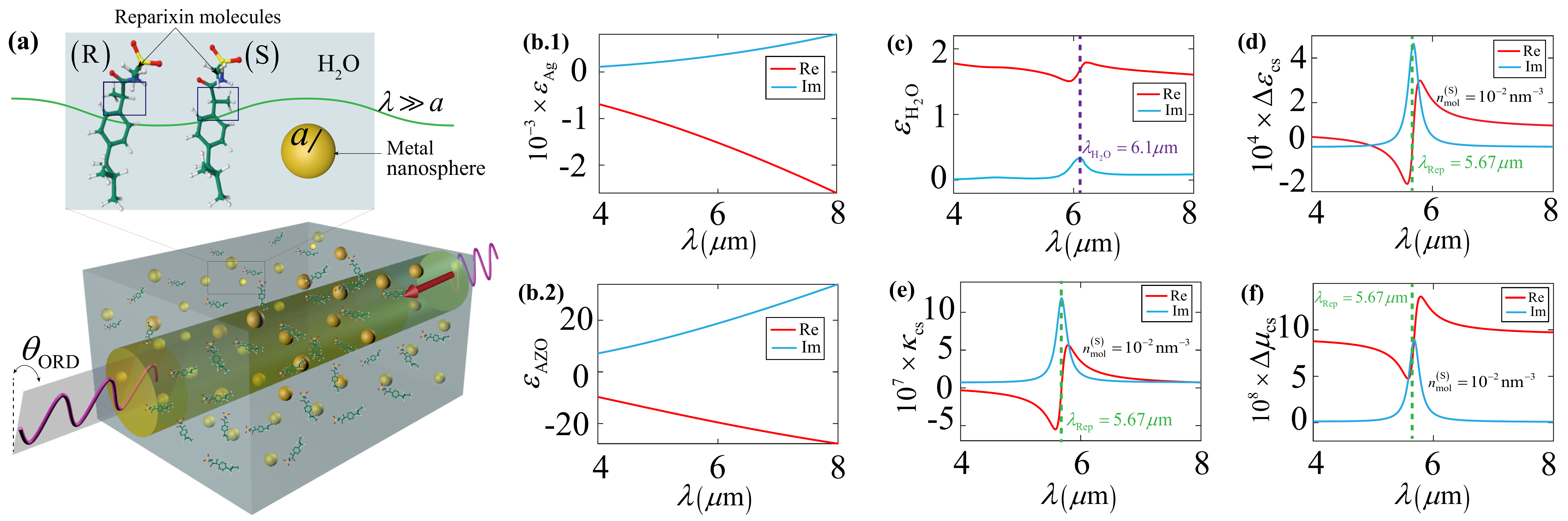}
		\caption{\textbf{(a)} Schematic representation of the proposed CEM composed of metallic NSs randomly dispersed in a solution of S and R reparixin enantiomers dissolved in water with to molecular density $n_{\rm mol}=n_{\rm mol}^{\rm (S)}+n_{\rm mol}^{\rm (R)}$. The vacuum wavelength of the impinging radiation $\lambda>>a,d$ is much larger than the NSs' radii $a$ and separations $d$.  \textbf{(b-f)} Dependence of \textbf{(b.1)} silver RDP $\epsilon_{\rm Ag}(\lambda)$ \cite{rakic2023},  \textbf{(b.2)} AZO RDP $\epsilon_{\rm AZO}(\lambda)$ \cite{shkondin2017}, \textbf{(c)} water RDP $\epsilon_{\rm H_2 O}(\lambda)$ \cite{laurens1999}, \textbf{(d)} reparixin RDP correction $\Delta \epsilon_{\rm cs}(\lambda)$ \cite{venturi2023}, \textbf{(e)} reparixin RMP correction $\Delta \mu_{\rm cs}(\lambda)$ \cite{venturi2023}, and \textbf{(f)} reparixin chiral parameter $\kappa_{\rm cs}(\lambda)$ \cite{venturi2023} over the impinging radiation vacuum wavelength $\lambda$. While reparixin RDP and RMP corrections do not depend over the individual enantiomer densities $n_{\rm mol}^{\rm (S,R)}$ but only on the total molecular density [$n_{\rm mol}=n_{\rm mol}^{\rm (S)}+n_{\rm mol}^{\rm (R)}=10^{-2} \rm nm^{-3}$ in \textbf{(c-f)}], the chiral parameter $\kappa_{\rm cs}(\lambda)$ depends over the enantiomeric density imbalance $\Delta n_{\rm mol}=n^{(\rm S)}_{\rm mol}-n^{(\rm R)}_{\rm mol}$, which in \textbf{(f)} is set to $\Delta n_{\rm mol} =n^{(\rm S)}_{\rm mol}=10^{-2} \rm nm^{-3}$.}
		\label{Fig1}
	\end{center}
\end{figure*}

\textit{Materials and settings}-- Fig. \ref{Fig1}(a) illustrates the considered CEM, composed of metallic NSs randomly dispersed in a chiral sample (CS). In our calculations we focus on an isotropic assembly of reparixin molecules dissolved in water ($\rm H_2 O + C_{11}H_{12}F_{3}NO_{6}S_{2}$), with a dilute molecular density  $n_{\rm mol}=n^{(\rm R)}_{\rm mol}+n^{(\rm S)}_{\rm mol} \simeq  10^{-2} \rm nm^{-3}$ (corresponding to a concentration of $\simeq$ 5 mg/ml), where $n^{(\rm R,S)}_{\rm mol}$ indicate the number densities of R and S enantiomers, respectively. We consider monochromatic optical excitation of the considered CEM by electric $\textbf{E}(\textbf{r},t)=\text{Re}\big[\textbf{E}_0(\textbf{r})e^{-i\omega t}\big]$ and magnetic $\textbf{H}(\textbf{r},t)=\text{Re}\big[\textbf{H}_0(\textbf{r})e^{-i\omega t}\big]$ fields with angular frequency $\omega=2\pi c/\lambda$, where $\lambda$ is the vacuum wavelength and $c$ is the speed of light in vacuum. The optical response of the metal is determined by the RDP $\epsilon_{\rm m}(\omega_0)$, which is depicted in  Figs. \ref{Fig1}(b.1,2) for (b.1) silver (Ag) and (b.2) Aluminium-doped zinc oxide (AZO). In the absence of metal nanoparticles, the monochromatic linear electromagnetic response of the isotropic CS is accounted for by unperturbed bi-anisotropic constitutive relations (BACRs) 
\begin{subequations}
\begin{align}
\textbf{D}_0^{\rm un}(\textbf{r})=\epsilon_0 \epsilon_{\rm cs}(\lambda) \textbf{E}_0(\textbf{r})-\frac{i}{c}\kappa_{\rm cs}(\lambda) \textbf{H}_0(\textbf{r}), \\
\textbf{B}_0^{\rm un}(\textbf{r})=\frac{i}{c}\kappa_{\rm cs}(\lambda) \textbf{E}_0(\textbf{r}) + \mu_0 \mu_{\rm cs}(\lambda)\textbf{H}_0(\textbf{r}), 
\end{align}
\end{subequations}
where $\textbf{D}_{\rm un}(\textbf{r},t)=\text{Re}\big[\textbf{D}_0^{\rm un}(\textbf{r})e^{-i\omega t}\big]$ and $\textbf{B}_{\rm un}(\textbf{r},t)=\text{Re}\big[\textbf{B}_0^{\rm un}(\textbf{r})e^{-i\omega t}\big]$ are the unperturbed electric displacement and magnetic induction vectors respectively, and $\epsilon_{\rm cs}(\lambda)$, $\mu_{\rm cs}(\lambda)$, and $\kappa_{\rm cs}(\lambda)$ are the wavelength-dependent macroscopic relative permittivity, permeability, and chirality parameter of the considered CS (aqueous reparixin). To calculate the BACR optical constants of the CS, we adopt a previously reported combined molecular dynamics simulation and quantum mechanical ensemble average approach \cite{venturi2023}, evaluating the macroscopic polarisation $\textbf{P}(\textbf{r},t)=\sum_{a=\text{R},\text{S}} n^{(a)}_{\text{mol}}\braket{\textbf{d}_a\left(\textbf{r},t\right)}$ and magnetisation $\textbf{M}(\textbf{r},t)=\sum_{a=\text{R},\text{S}} n^{(a)}_{\text{mol}}\braket{\textbf{m}_a\left(\textbf{r},t\right)}$ fields produced by reparixin molecules, where $\braket{\textbf{d}_a\left(\textbf{r},t\right)}$ and $\braket{\textbf{m}_a\left(\textbf{r},t\right)}$ are orientation-averaged electric and magnetic induced dipole moments of reparixin, respectively. Such polarisation and magnetisation fields enable the evaluation of the BACR parameters $\epsilon_{\rm cs}(\lambda)=\epsilon_{\rm H_2 O}(\lambda)+n_{\rm mol} A_{\rm ee}(\lambda)$, $\kappa_{\rm cs}=i\sum_{a=\text{R},\text{S}}n^{(a)}_{\text{mol}}A_{\rm em}^{(a)}(\lambda)$, and $\mu_{\rm cs}(\lambda)=1+n_{\rm mol} A_{\rm mm}(\lambda)$, where $A_{\rm ee}$, $A_{\rm mm}$, and $A_{\rm em}^{(a)}$ are orientation-averaged electric, magnetic, and magneto-electric (mixing) polarisabilities of every reparixin R,S enantiomer. Note that the chiral parameter of the CS is determined solely by the mixing polarisability, which flips sign for opposite enantiomers, $A_{\rm em}^{(\rm R)}=-A_{\rm em}^{(\rm S)}$, and in turn depends over the enantiomeric density imbalance $\Delta n_{\rm mol}=n^{(\rm S)}_{\rm mol}-n^{(\rm R)}_{\rm mol}$, yielding vanishing VOA for racemic mixtures where $\Delta n_{\rm mol}=0$. Note that $\epsilon_{\rm cs}(\lambda)$ accounts for both the dielectric permittivity of the solvent (water) $\epsilon_{\rm H_2 O}(\lambda)$ and the correction $\Delta \epsilon_{\rm cs}(\lambda)$ produced by reparixin. 
\begin{figure*}[t!]
	\centering
	\begin{center}
		\includegraphics[width=\textwidth ]{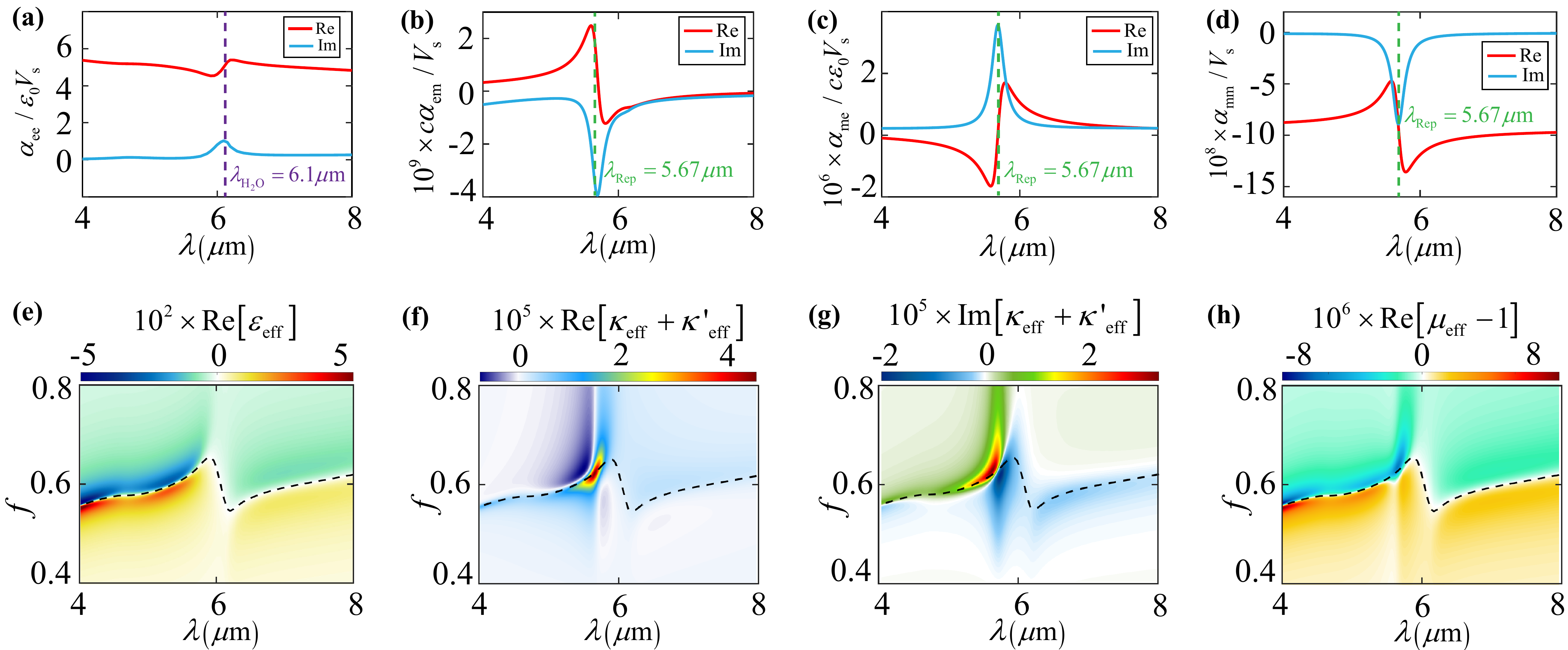}
		\caption{\textbf{(a-d)} Mid-IR wavelength-dependence of normalised {\bf (a)} electric $(\alpha_{\rm ee}/\epsilon_0 V_{\rm s})$, {\bf (b)} electric-magneto $(c \alpha_{\rm em}/V_{\rm s})$, {\bf (c)} magneto-electric $(\alpha_{\rm me}/c\epsilon_0 V_{\rm s})$, and {\bf (d)} magnetic $( \alpha_{\rm mm}/V_{\rm s})$ polarisabilities of a silver NS of volume $V_{\rm s}$ embedded in aqueous reparixin. \textbf{(e-h)} Effective mid-IR response of the bulk CEM (silver NSs$+$water$+$ reparixin) as function of NS filling fraction $f$ and impinging wavelength $\lambda$: \textbf{(e)} real part of effective RDP $\text{Re}[\epsilon_{\text{eff} }\left(f,\lambda\right)]$, \textbf{(f)} real and \textbf{(g)} imaginary parts the of $ \left(\kappa_{\rm eff}+\kappa_{\rm eff}'\right)$, and \textbf{(h)} real part of effective magnetic permeability $ \left(\mu_{\rm eff}-1\right)$. The dashed lines in (e-h) indicate the NZI condition $f_{\rm NZI}(\lambda)$ calculated from $\text{Re}[\epsilon_{\text{eff}}(f,\lambda)]=0$. In (e-h) the total molecular density is  $n_{\rm mol} =n^{(\rm S)}_{\rm mol}=10^{-2} \rm nm^{-3}$ and (f,g) illustrate results for a pure S reparixin solution with enantiomeric imbalance density $\Delta n_{\rm mol} =n^{(\rm S)}_{\rm mol}=n_{\rm mol}$.} 
		\label{Fig2}
	\end{center}
\end{figure*}
In Figs. \ref{Fig1}(c-f), we illustrate the dependence of  $\epsilon_{\rm H_2 O}$, $\Delta \epsilon_{\rm cs}$, $\kappa_{\rm cs}$, and $\mu_{\rm cs}$ over the impinging vacuum wavelength $\lambda$ for  pure reparixin S enantiomer dissolved in water with number density $n_{\rm mol}^{\rm (S)}=n_{\rm mol}=10^{-2}$ nm$^{-3}$. Note that, in the mid-infrared (mid-IR) range, the chiroptical response of reparixin is dominated by resonant vibrational transitions of water at $\lambda_{\rm H_2O} \simeq 6.10$ $\mu \rm m$ and reparixin at $\lambda_{\rm Rep} \simeq 5.67$ $\mu \rm m$. Due to the considered dilute molecular number density ($10^{-3} \rm nm^{-3}<n_{\rm mol}<10^{-1} \rm nm^{-3}$) of reparixin, the circular-polarisation (CP) averaged refractive index  $n(\lambda)=\text{Re}\big[\sqrt{\epsilon_{\rm cs}\mu_{\rm cs}}\big]\simeq\text{Re}\big[\sqrt{\epsilon_{\rm H_2O}}\big]$ and extinction coefficient $\beta_{\rm ext}=\text{Im}\big[\sqrt{\epsilon_{\rm cs}\mu_{\rm cs}}\big]\simeq\text{Im}\big[\sqrt{\epsilon_{\rm H_2O}}\big]$ are dominated by water. Indeed both the dielectric $\Delta \epsilon_{\rm cs}\simeq 10^{-4}$ and magnetic $\Delta \mu_{\rm cs}=\mu_{\rm cs}-1\simeq 10^{-8}$ corrections to the RDP and magnetic permeability, respectively, arising from reparixin are very weak, see Figs. \ref{Fig1}(d,f). Also note that the ORD $\propto \text{Re}[\kappa_{\rm cs}]l$ and CD $\propto \text{Im}[\kappa_{\rm cs}]l$ of the CS depend on real and imaginary parts of the chiral parameter, respectively, and on the CS length $l$, and thus are maximised around the reparixin vibrational resonance, see Fig. \ref{Fig1}(e).  
\begin{figure}[t!]
	\centering
	\begin{center}		\includegraphics[width=0.5\textwidth ]{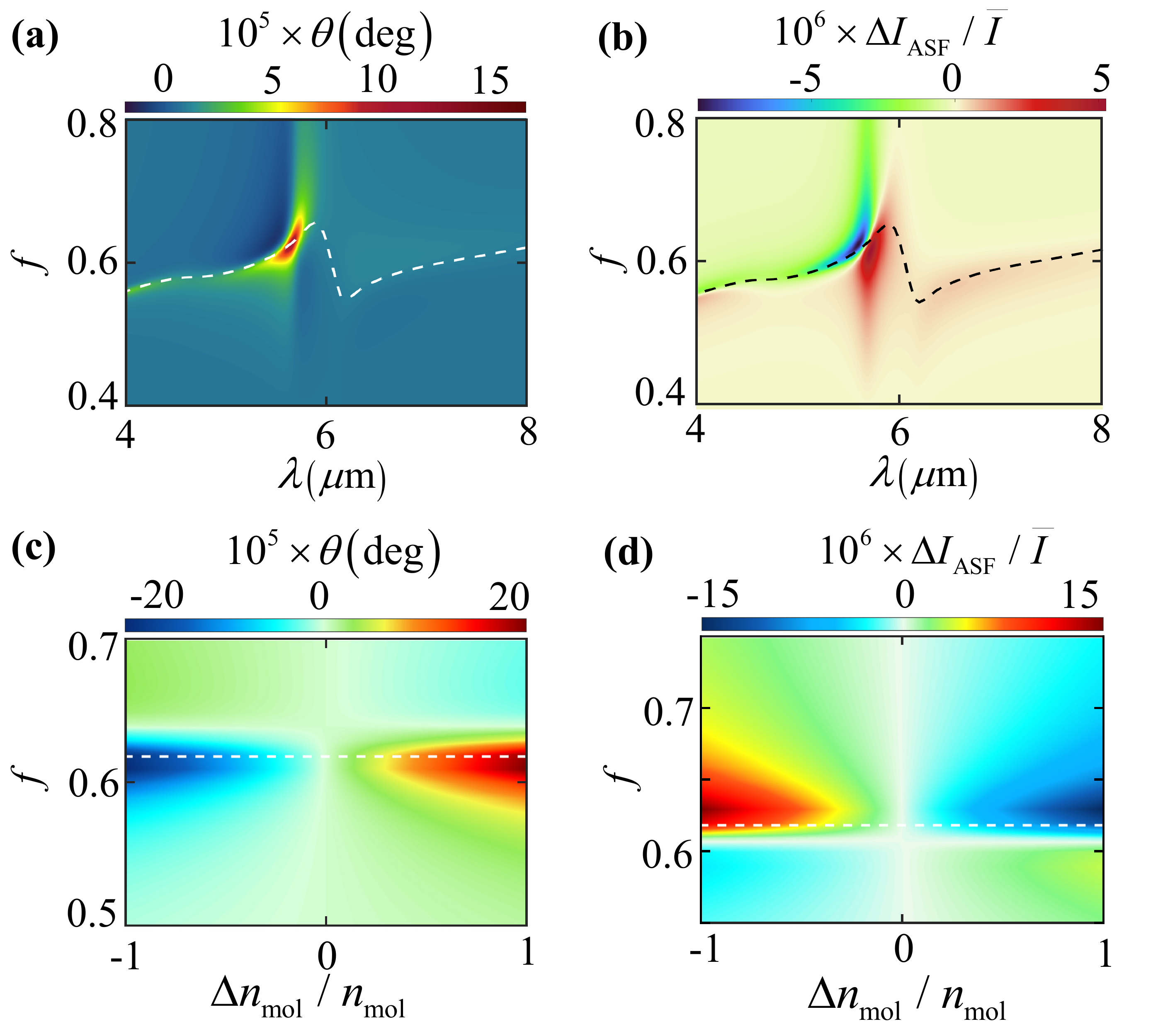}
		\caption{ Dependence of NZI-enhanced \textbf{(a)} ORD $\theta$ and \textbf{(b)} ASF $I_{\rm ASF}/\bar{I}$  over $f$ and $\lambda$ for a NZI-CEM length $l=100 \; \rm nm$ and $\Delta n_{\rm mol} =n^{(\rm S)}_{\rm mol}=10^{-2} \rm nm^{-3}$. Dependence of \textbf{(c)} ORD $\theta$ and \textbf{(d)} ASF over $f$ and $\Delta n_{\rm mol}/n_{\rm mol}$ for fixed $\lambda=5.6 \rm \; \mu m$, $l=100 \; \rm nm$, and $n_{\rm mol}=2\times 10^{-2} \rm nm^{-3}$. The dashed line in every plot indicates the NZI condition.} 
		\label{Fig3}
	\end{center}
\end{figure}
\textit{Effective mid-IR response}-- We evaluate the chiroptical response of the metallic NSs embedded in the CS host medium in the quasi-static limit $a \ll \lambda$, where $a$ is the NS radius, see sketch in Fig. \ref{Fig1}(a), providing only electric/magnetic induced dipoles and neglecting higher-order induced multipoles. Indeed, for such small NSs retardation effects inside the NS are negligible, as the induced polarisation charges follow the external field adiabatically over time, and the field distribution is practically uniform within the NS. In this approximation, one can introduce electric (E) and magnetic (M) scalar potentials inside ($\Phi^{\rm E, M}_{\rm in}$) and outside ($\Phi^{\rm E, M}_{\rm out}$) the NS. Such potentials satisfy the Laplace equation with solutions $\Phi^{\rm E, M}_{\rm in}=\sum_{l'=0}^{\infty}\sum_{m'=-l'}^{l'}\mathcal{A}^{\rm E, M}_{l'm'}r^{l'} Y_{l'm'}(\theta,\phi)$ and $\Phi^{\rm E, M}_{\rm out}=\sum_{l'=0}^{\infty}\sum_{m'=-l'}^{l'}\left(\mathcal{C}^{\rm E, M}_{l'm'}r^{l'}+\mathcal{D}^{\rm E, M}_{l'm'}r^{-(l'+1)}\right) Y_{l'm'}(\theta,\phi)$, where $Y_{l'm'}(\theta,\phi)$ are spherical harmonics, and the coefficients $\mathcal{A}^{\rm E, M}_{l'm'}$, $\mathcal{C}^{\rm E, M}_{l'm'}$, and $\mathcal{D}^{\rm E, M}_{l'm'}$ are calculated by applying the electromagnetic boundary conditions at the interface between the metal NS and the outside CS, see the Appendix for further details. Retaining only the dipolar $(l'=1)$ contributions, the mid-IR response of the NS reduces to point-like electric/magnetic dipoles described by the dynamic polarisability tensor $\hat{\alpha}$, which linearly maps the impinging fields to the induced electric and magnetic dipole moments $\textbf{p}$ and $\textbf{m}$ respectively,
\begin{equation} \label{pminduced} 
\begin{bmatrix}
\textbf{p}\\
\textbf{m}
\end{bmatrix}=\begin{bmatrix}
\alpha_{\rm ee} & i\alpha_{\rm em}\\
-i\alpha_{\rm me} & \alpha_{\rm mm}
\end{bmatrix} \begin{bmatrix}
\textbf{E}_0\\
\textbf{H}_0
\end{bmatrix}.
\end{equation}
In the expression above,  $\alpha_{\rm ee}=3V_s \gamma^{-1} \epsilon_0  \epsilon_{\rm cs} \big[(\epsilon_{\rm m}-\epsilon_{\rm cs})(2\mu_{\rm cs}+1) + 2\kappa_{\rm cs}^2\big]$, $\alpha_{\rm mm}=-3V_s\gamma^{-1} \big[(\epsilon_{\rm m}+2\epsilon_{\rm cs})(\mu_{\rm cs}-1)-2\kappa_{\rm cs}^2\big]$ are the electric and magnetic co-polarisability tensor elements, and  $\alpha_{\rm em}= 9V_s\left(c\gamma\right)^{-1} \kappa_{\rm cs} \epsilon_{\rm cs}$, $\alpha_{\rm me}= 9V_s\gamma^{-1} c \epsilon_0 \kappa_{\rm cs}  \epsilon_{\rm m}$ are the mixing polarisability tensor elements, where $\gamma=(\epsilon_{\rm m}+2\epsilon_{\rm cs}) (2\mu_{\rm cs}+1)-4\kappa_{\rm cs}^2$, and $V_s=(4/3)\pi a^3$ is the NS volume. Figs. \ref{Fig2}(a-d) illustrate the wavelength-dependence of $\alpha_{\rm ee},\alpha_{\rm me},\alpha_{\rm em},\alpha_{\rm mm}$ for a silver NS embedded in the considered CS, which are resonant at the reparixin [$\alpha_{\rm me},\alpha_{\rm em},\alpha_{\rm mm}$, see Figs. \ref{Fig2}(b-d)] and water [$\alpha_{\rm ee}$, see Fig. \ref{Fig2}(a)] vibrational transitions, $\lambda_{\rm Rep} \simeq 5.67$ $\mu$m and $\lambda_{\rm H_2O} \simeq 6.10$ $\mu$m, respectively. It is worth emphasizing that, because the considered wavelength range is highly detuned from the NS localised surface plasmon resonance, these results are practically independent over the noble metal of choice, observing almost identical behaviours in all the considered cases. By analogy with the unperturbed CS, we anticipate that strong VOA produced by the considered plasmonic NSs arises from the interplay between the electric and magnetic induced dipoles, i.e., from the mixing polarisabilities $\alpha_{\rm em,me}$. To calculate the effective CEM response, we adopt the Maxwell-Garnett (MG) homogenisation approach in the limit $d\simeq n_{\rm mol}^{1/3}\simeq 10$ nm $<<\lambda$, taking into account the local field corrections \cite{merkel2026}. Considering a macroscopic volume with NS number density $n$, we obtain the effective BACRs
\begin{subequations}
\begin{align}
\textbf{D}_0(\textbf{r})=\epsilon_0 \epsilon_{\rm eff}(\lambda,f) \textbf{E}_0(\textbf{r})-\frac{i}{c}\kappa_{\rm eff}(\lambda,f)\textbf{H}_0(\textbf{r}),  \\ 
\textbf{B}_0(\textbf{r})=\frac{i}{c}\kappa_{\rm eff}'(\lambda,f) \textbf{E}_0(\textbf{r}) + \mu_0 \mu_{\rm eff}(\lambda,f) \textbf{H}_0(\textbf{r}),
\end{align}
\end{subequations}
where $\epsilon_{\text{eff}}= \Gamma^{-1}\big[\epsilon_{\text{cs}} +(2n/3\epsilon_0)(\alpha_{\rm ee}+c\epsilon_0\kappa_{\rm cs} \alpha_{\rm em}+\epsilon_0\epsilon_{\text{cs}}\alpha_{\rm mm}+(4n^2/9\epsilon_0)(\alpha_{\rm ee}\alpha_{\rm mm}-\alpha_{\rm em}\alpha_{\rm me})\big]$,  $\kappa_{\text{eff}}=\Gamma^{-1}\big[\kappa_{\text{cs}}+(2n/3)(\kappa_{\rm cs}\alpha_{\rm mm}+c\mu_{\rm cs} \alpha_{\rm em})-(5n/3)c\alpha_{\rm em}\big]$,  $\kappa'_{\text{eff}}=\Gamma^{-1}\big[\kappa_{\text{cs}}-(n/3)c\mu_0(\epsilon_{\rm cs}+2)\alpha_{\rm me}-(n/3)\mu_0 c^2 \kappa_{\rm cs} \alpha_{\rm ee}\big]$,  $\mu_{\text{eff}}=\Gamma^{-1}\big[\mu_{\text{cs}}+(2n/3)\alpha_{\rm mm}-(n/3\epsilon_0)\mu_{\rm cs}\alpha_{\rm ee} + (5n^2/9\epsilon_0)(\alpha_{\rm me} \alpha_{\rm em}-\alpha_{\rm mm}\alpha_{\rm ee})-(n^2/3\epsilon_0 c)\kappa_{\rm cs} \alpha_{\rm me} \alpha_{\rm mm}\big]$, and $\Gamma = \big[1+(2n/3)\alpha_{\rm mm}-(n/3\epsilon_0)\alpha_{\rm ee}-(2n^2/9\epsilon_0)(\alpha_{\rm ee}\alpha_{\rm mm}-\alpha_{\rm em}\alpha_{\rm me})\big]$, see the Appendix. 

\begin{figure}[t]
	\centering
	\begin{center}		\includegraphics[width=0.5\textwidth ]{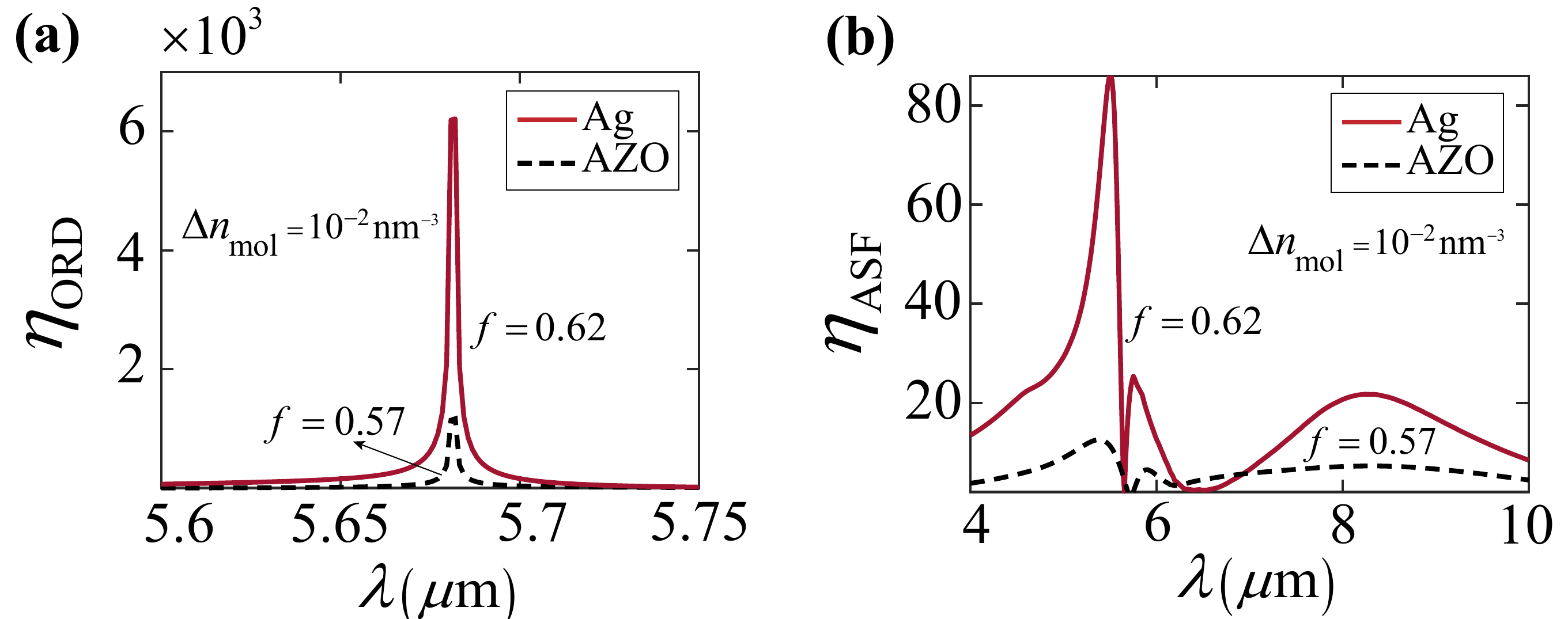}
		\caption{\textbf{(a)} ORD $\eta_{\rm ORD}(\lambda)$ and \textbf{(b)} ASF $\eta_{\rm ASF}(\lambda)$ enhancement factors for Ag (red solid lines) and AZO (black dashed lines) based CEMs with $f=0.62$ and $f=0.57$, respectively. In all plots $\Delta n_{\rm mol}=n^{(S)}_{\rm mol}= 10^{-2} \rm nm^{-3}$ and $l=100 \; \rm nm$. } 
		\label{Fig4}
	\end{center}
\end{figure}

Note that, owing to the polarisabilities proportionality to the NS volume, the above expressions depend only over the NS filling fraction $f=(4/3)\pi a^3 n$. Figs. \ref{Fig2}(e-h) illustrate the dependence of $\epsilon_{\rm eff}(\lambda,f)$, $\mu_{\rm eff}(\lambda,f)$ and $\kappa_{\rm eff}(\lambda,f)+\kappa'_{\rm eff}(\lambda,f)$ over the excitation wavelength $\lambda$ and the filling fraction $f$ of Ag NSs. Note that the NZI condition dictated by $\text{Re}[\epsilon_{\rm eff}(f,\lambda)]=0$ is achieved for certain combinations of $f$ and $\lambda$ defining the NZI curve $f_{\rm ENZ}(\lambda)$, indicated by the dashed lines in Figs.  \ref{Fig2}(e-h). To quantify the VCD differential absorption of left-circularly polarised (LCP) and right-circularly polarised (RCP) light, we define the asymmetry factor (ASF) as $\Delta I_{\rm ASF}/\bar{I}$, where $\Delta I_{\rm ASF}=I^{(+)}_{\rm T}-I^{(-)}_{\rm T}$, $\bar{I}=(I^{(+)}_{\rm T}+I^{(-)}_{\rm T})/2$, and $I_{\rm T}^{(\pm 1)}$ is the transmitted intensity upon $s=\pm 1$ excitation by RCP $(s=+1)$ and LCP $(s=-1)$. By inserting Eqs. (3a,b) into macroscopic Maxwell's equations, one finds that RCP/LCP plane waves are eigenvectors with complex refractive indexes $n_{\pm}= (1/2)\big[\mp (\kappa_{\rm eff} + \kappa_{\rm eff}^{'})+ \sqrt{(\kappa_{\rm eff} - \kappa_{\rm eff}^{'})^2 + 4\epsilon_{\rm eff} \mu_{\rm eff} }\big]$ and  corresponding wavenumbers $\beta_{\pm}=2\pi n_{\pm}/\lambda$. In turn, circularly polarised light with input intensity $I_0$ maintains its polarisation state and is transmitted with an intensity $I^{(s)}_{\rm T}=I_0 e^{-2{\rm Im}\beta_s l}$, providing $\Delta I_{\rm ASF}/\bar{I}=-2 \text{tanh}\big\{2\pi l\lambda^{-1}\text{Im}[n_+-n_-]\big\}$. Conversely, for linearly polarised excitation, the polarisation direction rotates over progagation with a final ORD angle $\theta(\lambda)=\pi l \lambda^{-1}\text{Re}[n_+(\lambda)-n_-(\lambda)]$. Figs. \ref{Fig3}(a,b) illustrate the dependence of VOA (a) ORD and (b) ASF for a silver-based CEM with $l=100 \; \rm nm$ (and picolitre volume $\simeq 25$ pl assuming lateral dimensions of $\simeq500$ $\mu$m), indicating that both ORD and VCD are highly enhanced at NZI conditions. Figs. \ref{Fig3}(c,d) further illustrate the dependence of (c) ORD and (d) ASF over $\Delta n_{\rm mol}/n_{\rm mol}$ and $f$ for fixed excitation wavelength $\lambda=5.6 \; \rm \mu m$. We emphasize that the choice of noble metals other than silver provides similar results in the mid-IR wavelength range, which is highly detuned from the plasma frequency. While in principle the adoption of transparent conductors with plasma frequencies in the IR, e.g., AZO, see Fig. \ref{Fig1}(b.2), can provide a further knob to modulate the enhancement factor, we also observe similar results with AZO and ITO. For the sake of comparison, in Figs. \ref{Fig4}(a,b) we depict the (a) ORD and (b) ASF enhancement factors $\eta_{\rm ORD}=\theta/\theta_{\rm un}$ and $\eta_{\rm ASF}=\Delta I_{\rm ASF}/\Delta I_{\rm ASF}^{\rm (un)}$ for disordered CEMs composed of Ag and AZO at the fixed filling fractions $f=0.62$ (Ag) and  $f=0.57$ (AZO), respectively, where $\theta_{\rm un}$ and $\Delta I_{\rm ASF}^{\rm (un)}$ are the ORD angle and ASF for vanishing filling fraction ($f=0$), i.e., in the absence of the NSs. Note that the maximum enhancement attained at NZI conditions is the one of Ag (${\rm max} ~ \eta_{\rm ORD} \simeq 10^3$ and ${\rm max}~ \eta_{\rm ASF} \simeq 100$), while AZO provides smaller NZI enhancement owing to the larger field penetration within the NS. Note that, in spite of the distinct conductor choice, the filling fractions required to attain NZI conditions are really similar, thus making our proposed platform robust over metallic behavoiur, including oxidation and imperfections.

\textit{Conclusions}---In conclusion, we find that disordered CEMs consisting of randomly dispersed metal-based NSs embedded within an optically active medium greatly enhance VOA thanks to NZI-induced superchirality. Our first principle calculations indicate that NZI conditions, attained in the mid-IR by adjusting the NSs filling fraction, enhance ORD by a factor $\simeq 6\times 10^3$ and VCD ASF by a factor $\simeq 10^2$. We further observe that silver-based NSs are the best candidates maximising VOA thanks to high reflectivity in the mid-IR, reducing field penetration within the NS with respect to transparent conductors like AZO.  Our results shed light on the possibilities offered by nanophotonics for detecting chirality at the pl-volume level, with potentially disruptive applications in quantum chemistry and nanomedicine.

This work has been partially funded by the European Union - NextGenerationEU under the Italian Ministry of University and Research (MUR) National Innovation Ecosystem grant ECS00000041 - VITALITY - CUP E13C22001060006. This work has been supported by the European Union under grant agreement No 101046424. Views and opinions expressed are however those of the author(s) only and do not necessarily reflect those of the European Union or the European Innovation Council. Neither the European Union nor the European Innovation Council can be held responsible for them.

The authors acknowledge fruitful discussions with Massimiliano Aschi. Jens Biegert, Francesco Tani, Patrice Genevet, Samira Khadir, Remi Colom, Michele Dipalo, Giovanni Melle, Sotirios Christodoulou, and Anna Maria Cimini.

\onecolumngrid
\appendix

\section*{Appendix: Supplementary Information}

Here, we derive the electric, magnetic, and magnetoelectric polarisability expressions (in Eq. 2 of the main text) for a metal nanosphere embedded in a chiral host medium. Next, we derive the effective optical constants of the metamaterial from Maxwell-Garnett theory. 

\section{Polarisabilities of a metal sphere embedded in a chiral host medium}
\begin{figure}[H]
	\centering
	\begin{center}
		\includegraphics[scale=0.1]{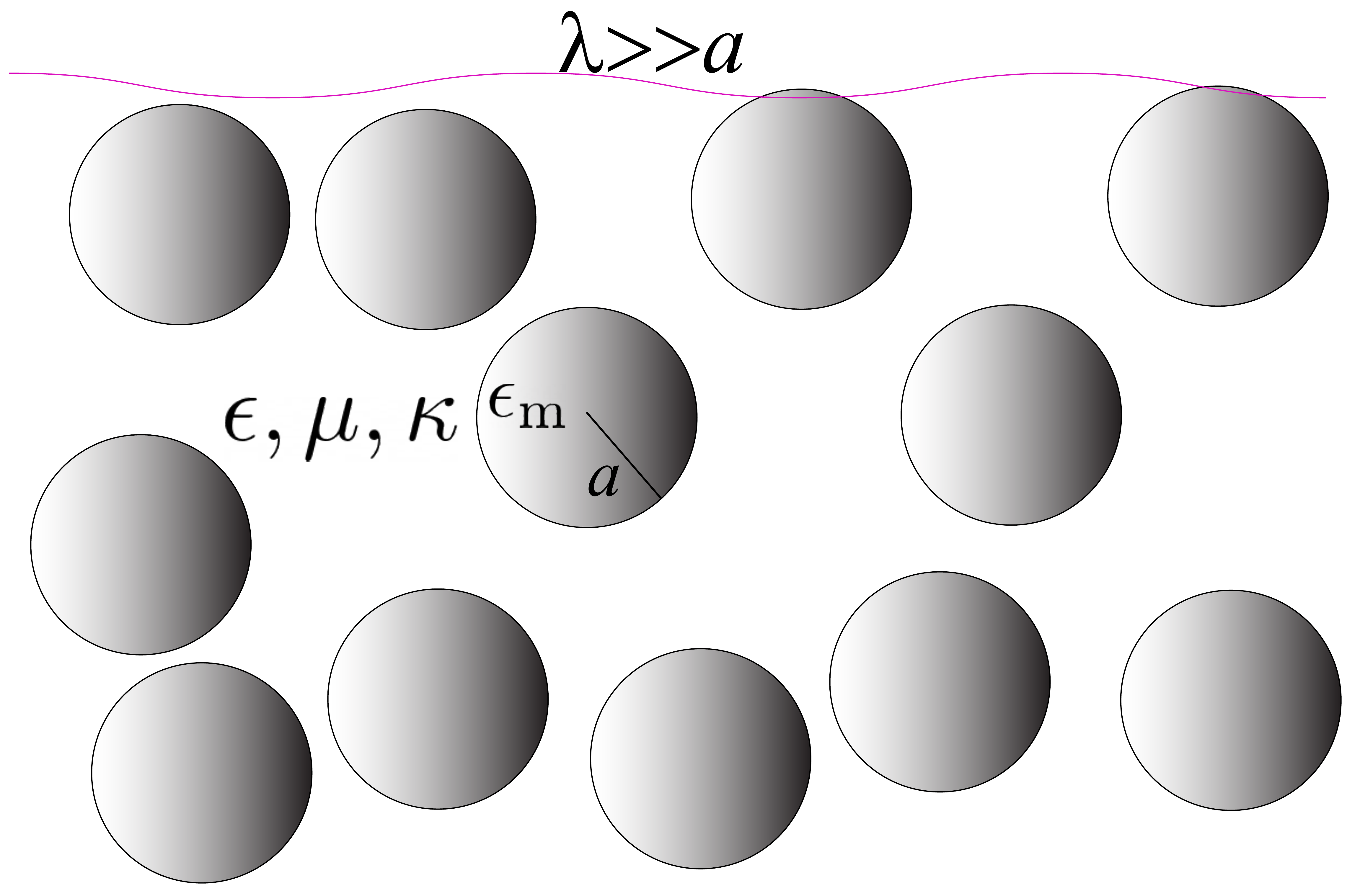}
		\captionsetup{labelformat=empty}
		\caption{FIG1. Schematic illustration of metal spheres of radius $a$ and permittivity $\epsilon_{\rm m}$ embedded in a chiral background characterized by $\epsilon$, $\mu$, $\kappa$. } 
		\label{fig1si}
	\end{center}
\end{figure}
Consider a metal nanosphere (NS) of radius $a$ and dielectric permittivity $\epsilon_{\rm m}$ embedded in a chiral host (CH) characterized by permittivity $\epsilon$, permeability $\mu$ and chiral parameter $\kappa$. The constitutive relations inside the metal sphere are
\begin{subequations}
	\begin{align}
		\label{crp_a}
		\textbf{D}_{\rm in} &= \epsilon_0 \epsilon_{\rm m} \textbf{E}_{\rm in},\\	\label{crp_b}
		\textbf{B}_{\rm in} &= \mu_0 \textbf{H}_{\rm in},
	\end{align}
\end{subequations}
while for the outer medium are
\begin{subequations}
	\begin{align}
		\label{crm_a}
		\textbf{D}_{\rm out} &= \epsilon_0 \epsilon \textbf{E}_{\rm out}- i\frac{\kappa}{c} \textbf{H}_{\rm out},\\			
		\label{crm_b}
		\textbf{B}_{\rm out} &= \mu_0 \mu \textbf{H}_{\rm out}+ i \frac{\kappa}{c} \textbf{E}_{\rm out}.
	\end{align}
\end{subequations}
In the electrostatic limit $(\lambda \gg a)$, we can define scalar potentials $\Phi_{\rm in}^{(\rm E, M)}$ and $\Phi_{\rm out}^{(\rm E, M)}$ for the electric and magnetic field vectors 
\begin{subequations}
	\begin{align}
		\label{phin}
		\begin{cases}
			\textbf{E}_{\rm in} &= - \nabla \Phi^{(\rm E)}_{\rm in} \\	
			\textbf{H}_{\rm in} &= - \nabla \Phi^{(\rm M)}_{\rm in}
		\end{cases}, \\
		\label{phiout}
		\begin{cases}
			\textbf{E}_{\rm out} &= - \nabla \Phi^{(\rm E)}_{\rm out} \\	
			\textbf{H}_{\rm out} &= - \nabla \Phi^{(\rm M)}_{\rm out}
		\end{cases}. 
	\end{align}
\end{subequations}
Such potentials satisfy the Laplace equations inside and outside the sphere
\begin{subequations}
	\begin{align}
		\nabla^2 \Phi_{\text{in}}^{(\rm E)} &= 0, \quad r \leq a, \\
		\nabla^2 \Phi_{\text{in}}^{(\rm M)} &= 0, \quad r \leq a, \\
		\nabla^2 \Phi_{\text{out}}^{(\rm E)} &= 0, \quad r \geq a, \\
		\nabla^2 \Phi_{\text{out}}^{(\rm M)} &= 0, \quad r \geq a.
	\end{align}
	\label{eq:13}
\end{subequations}
with solutions given by
\begin{subequations}
	\begin{align}
		\Phi_{\text{in}}^{(\rm E)} &= \sum_{l=0}^\infty \sum_{m=-l}^l \mathcal{A}_{lm}^{(\rm E)} r^l Y_{lm}(\theta, \phi), \quad r \leq a, \\
		\Phi_{\text{in}}^{(\rm M)} &= \sum_{l=0}^\infty \sum_{m=-l}^l \mathcal{A}_{lm}^{(\rm M)} r^{l}  Y_{lm}(\theta, \phi), \quad r \leq a, \\
		\Phi_{\text{out}}^{(\rm E)} &= \sum_{l=0}^\infty \sum_{m=-l}^l \big(\mathcal{C}_{lm}^{(\rm E)} r^l + \mathcal{D}_{lm}^{(\rm E)} r^{-(l+1)}\big)  Y_{lm}(\theta, \phi), \quad r \geq a, \\
		\Phi_{\text{out}}^{(\rm M)} &= \sum_{l=0}^\infty \sum_{m=-l}^l \big(\mathcal{C}_{lm}^{(\rm M)} r^l + \mathcal{D}_{lm}^{(\rm M)} r^{-(l+1)}\big)  Y_{lm}(\theta, \phi), \quad r \geq a,
	\end{align}
	\label{eq:15}
\end{subequations}
where $Y_{lm}(\theta,\phi)$ are spherical harmonics. The coefficients $\mathcal{A}^{(\rm E, M)}_{lm}$, $\mathcal{C}^{(\rm E, M)}_{lm}$, and $\mathcal{D}^{(\rm E, M)}_{lm}$ are to be determined from the boundary conditions at the sphere's surface $r=a$  
\begin{subequations}
	\begin{align}
		\label{Ebc}
		\begin{cases}
			\text{D}_{\rm out}^{\bot} (r=a) &= \text{D}_{\rm in}^{\bot}(r=a) \\	
			\text{E}_{\rm out}^{||}(r=a) &= \text{E}_{\rm in}^{||}(r=a) \\
			\text{B}_{\rm out}^{\bot} (r=a) &= \text{B}_{\rm in}^{\bot}(r=a)\\
			\text{H}_{\rm out}^{||}(r=a) &= \text{H}_{\rm in}^{||}(r=a)
		\end{cases}
	\end{align}
\end{subequations}
and from the far field ($r \rightarrow \infty$) potentials
\begin{subequations}
	\begin{align}
		\label{phinf}
		\begin{cases}
			\Phi_{\rm E, out} (r \rightarrow \infty) &= -\textbf{E}_0 \cdot \boldsymbol{r} = - \text{E}_{0x} r \text{sin} \theta\text{cos}\phi - \text{E}_{0y} r \text{sin}\theta \text{sin}\phi - \text{E}_{0z} r \text{cos}\theta \\	
			\Phi_{\rm M, out} (r \rightarrow \infty) &= -\textbf{H}_0 \cdot \boldsymbol{r} = - \text{H}_{0x} r \text{sin}\theta\text{cos}\phi - \text{H}_{0y} r \text{sin}\theta \text{sin}\phi - \text{H}_{0z} r \text{cos}\theta
		\end{cases},
	\end{align}
\end{subequations}
which in terms of spherical harmonics become
\begin{subequations}
	\begin{align}
		\label{phinf1}
		\begin{cases}
			\Phi_{\rm E, out} (r \rightarrow \infty) &= \sqrt{\frac{2\pi}{3}} (\text{E}_{0x}-i\text{E}_{0y}) r Y_{1,1} - \sqrt{\frac{2\pi}{3}} (\text{E}_{0x}+i\text{E}_{0y}) r Y_{1,-1} -\sqrt{\frac{4\pi}{3}}  \text{E}_{0z} r Y_{1,0}\\	
			\Phi_{\rm M, out} (r \rightarrow \infty) &= \sqrt{\frac{2\pi}{3}} (\text{H}_{0x}-i\text{H}_{0y}) r Y_{1,1} - \sqrt{\frac{2\pi}{3}} (\text{H}_{0x}+i\text{H}_{0y}) r Y_{1,-1} -\sqrt{\frac{4\pi}{3}}  \text{H}_{0z} r Y_{1,0}
		\end{cases}.
	\end{align}
\end{subequations}
Taking the limit $r \rightarrow \infty$ in Eqs. \eqref{eq:15} and comparing with Eq. \eqref{phinf1} we obtain
\begin{subequations}
	\begin{align}  
		\begin{cases}
			\mathcal{C}^{(\rm E)}_{1,1}  &= \sqrt{\frac{2\pi}{3}} (\text{E}_{0x}-i\text{E}_{0y}) \\ 
			\mathcal{C}^{(\rm E)}_{1,-1}  &= -\sqrt{\frac{2\pi}{3}} (\text{E}_{0x}+i\text{E}_{0y}) \\ 
			\mathcal{C}^{(\rm E)}_{1,0}  &= -\sqrt{\frac{4\pi}{3}} \text{E}_{0z} 
			\label{clme}
		\end{cases}.
	\end{align}
\end{subequations}
\begin{subequations}
	\begin{align}  
		\begin{cases}
			\mathcal{C}^{(\rm M)}_{1,1}  &= ic\mu_0\sqrt{\frac{2\pi \mu}{3\epsilon}} (\text{H}_{0x}-i\text{H}_{0y}) \\ 
			\mathcal{C}^{(\rm M)}_{1,-1}  &= -ic\mu_0\sqrt{\frac{2\pi \mu}{3\epsilon}} (\text{H}_{0x}+i\text{H}_{0y}) \\ 
			\mathcal{C}^{(\rm M)}_{1,0}  &= -ic\mu_0\sqrt{\frac{4\pi \mu}{3\epsilon}} \text{H}_{0z} 
			\label{clmm}
		\end{cases}.
	\end{align}
\end{subequations}
In terms of the modified potentials, the boundary conditions in Eqs. \eqref{Ebc} become
\begin{eqnarray}
	\epsilon_0 \epsilon  \frac{\partial \Phi^{(\rm E)}_{\rm out}}{\partial r} \bigg\rvert_{r=a} - i\frac{\kappa}{c}  \frac{\partial \Phi^{(\rm M)}_{\rm out}}{\partial r}\bigg\rvert_{r=a}  &=& \epsilon_0 \epsilon_{\rm m}  \frac{\partial \Phi^{(\rm E)}_{\rm in}}{\partial r} \bigg\rvert_{r=a},  \\	
	\frac{\partial \Phi^{(\rm E)}_{\rm out}}{\partial \theta}\bigg\rvert_{r=a} &=&  \frac{\partial \Phi^{(\rm E)}_{\rm in}}{\partial \theta}\bigg\rvert_{r=a}, \\
	\mu_0 \mu  \frac{\partial \Phi^{(\rm M)}_{\rm out}}{\partial r}\bigg\rvert_{r=a} + i \frac{\kappa}{c} \frac{\partial \Phi^{(\rm E)}_{\rm out}}{\partial r}\bigg\rvert_{r=a}&=&\mu_0 \frac{\partial \Phi^{(\rm M)}_{\rm in}}{\partial r}\bigg\rvert_{r=a},  \\ 
	\frac{\partial \Phi^{(\rm M)}_{\rm out}}{\partial r}\bigg\rvert_{r=a}&=& \frac{\partial \Phi^{(\rm M)}_{\rm in}}{\partial r}\bigg\rvert_{r=a}.
\end{eqnarray}
Substituting the series expansion into the above equations and using the orthonormality of spherical harmonics we get a set of algebraic equations involving the coefficients $\mathcal{A}_{lm}, \mathcal{C}_{lm}$, and $\mathcal{D}_{lm}$. Eliminating the known coefficients $\mathcal{C}_{lm}$ from Eqs. \ref{clme}, \ref{clmm} we solve for the unknown $\mathcal{A}_{lm}$ and $\mathcal{D}_{lm}$, obtaining the only non-zero terms for $l=1$ and $m=-1,0,1$ 
\begin{subequations}
	\begin{align}  
		\begin{cases}
			\mathcal{D}_{1,0}^{(\rm E)}   &= a^3 \sqrt{\frac{4\pi}{3}} \left\{ \frac{[(\epsilon_{\rm m}-\epsilon)(2\mu+1)+2\kappa^2]\text{E}_{0z} + 3i c \mu_0 \kappa\text{H}_{0z}}{(\epsilon_{\rm m}+2\epsilon)(2\mu+1)-4\kappa^2} \right\} \\ \\
			\mathcal{D}_{1,0}^{(\rm M)} &= -ia^3 \sqrt{\frac{4\pi \mu}{3\epsilon}} \left\{ \frac{[c\mu_0(\epsilon_{\rm m}+2\epsilon)(\mu-1)-2\kappa^2]\text{H}_{0z} + 3i\kappa  \epsilon_{\rm m}\text{E}_{0z}}{(\epsilon_{\rm m}+2\epsilon)(2\mu+1)-4\kappa^2} \right\} 
			\label{dem10}
		\end{cases}.
	\end{align}
\end{subequations}
\begin{subequations}
	\begin{align}  
		\begin{cases}
			\mathcal{D}_{1,1}^{(\rm E)} &= -a^3 \sqrt{\frac{2\pi}{3}} \left\{ \frac{[(\epsilon_{\rm m}-\epsilon)(2\mu+1)+2\kappa^2](\text{E}_{0x}-i\text{E}_{0y}) + 3i c \mu_0 \kappa(\text{H}_{0x}-i\text{H}_{0y})}{(\epsilon_{\rm m}+2\epsilon)(2\mu+1)-4\kappa^2} \right\} \\ \\
			\mathcal{D}_{1,1}^{(\rm M)}  &= ia^3 \sqrt{\frac{2\pi \mu}{3\epsilon}} \left\{ \frac{[c\mu_0(\epsilon_{\rm m}+2\epsilon)(\mu-1)-2\kappa^2](\text{H}_{0x}-i\text{H}_{0y}) + 3i\kappa \epsilon_{\rm m}  (\text{E}_{0x}-i\text{E}_{0y})}{(\epsilon_{\rm m}+2\epsilon)(2\mu+1)-4\kappa^2} \right\} 
			\label{dem11}
		\end{cases}.
	\end{align}
\end{subequations}
\begin{subequations}
	\begin{align}  
		\begin{cases}
			\mathcal{D}_{1,-1}^{(\rm E)}  &= a^3 \sqrt{\frac{2\pi}{3}} \left\{ \frac{[(\epsilon_{\rm m}-\epsilon)(2\mu+1)+2\kappa^2](\text{E}_{0x}+i\text{E}_{0y}) + 3 i c  \mu_0  \kappa (\text{H}_{0x}+i\text{H}_{0y})}{(\epsilon_{\rm m}+2\epsilon)(2\mu+1)-4\kappa^2} \right\} \\ \\
			\mathcal{D}_{1,-1}^{(\rm M)} &= -ia^3 \sqrt{\frac{2\pi \mu}{3\epsilon}} \left\{ \frac{[c\mu_0(\epsilon_{\rm m}+2\epsilon)(\mu-1)-2\kappa^2](\text{H}_{0x}+i\text{H}_{0y}) + 3i  \kappa  \epsilon_{\rm m} (\text{E}_{0x}+i\text{E}_{0y})}{(\epsilon_{\rm m}+2\epsilon)(2\mu+1)-4\kappa^2} \right\} 
			\label{dem1m1}
		\end{cases}.
	\end{align}
\end{subequations}
The induced dipole moments $\textbf{p}$ and $\textbf{m}$ can be calculated from the dipole contribution ($l=1; m=-1,0,1$) to the potentials outside the sphere
\begin{subequations}
	\begin{align}
		\begin{cases}
			\Phi_{\rm E,out}|_{\text{dipole}} &=   \big[ \mathcal{D}_{1,0}^{(\rm E)}Y_{1,0} + \mathcal{D}_{1,1}^{(\rm E)} Y_{1,1} + \mathcal{D}_{1,-1}^{(\rm E)} Y_{1,-1}\big] \frac{1}{r^2} \equiv \frac{1}{4\pi \epsilon_0 \epsilon} \frac{\textbf{p} \cdot \hat{\boldsymbol{r}}}{r^2}\\	\\ \nonumber
			\Phi_{\rm M,out}|_{\text{dipole}} &= \big[ \mathcal{D}_{1,0}^{(\rm M)}Y_{1,0} + \mathcal{D}_{1,1}^{(\rm M)} Y_{1,1} + \mathcal{D}_{1,-1}^{(\rm M)} Y_{1,-1}\big] \frac{1}{r^2}  \equiv \frac{1}{4\pi } \frac{\textbf{m} \cdot \hat{\boldsymbol{r}}}{r^2}
		\end{cases} .
	\end{align}
\end{subequations}
After simplification, we obtain
\begin{equation} \label{pem1}
	\begin{bmatrix}
		\textbf{p}\\
		\textbf{m}
	\end{bmatrix} = \begin{bmatrix}
		\rttensortwo{\alpha}_{ee} & i\rttensortwo{\alpha}_{em} \\
		-i\rttensortwo{\alpha}_{me} & \rttensortwo{\alpha}_{mm}
	\end{bmatrix} \begin{bmatrix}
		\textbf{E}_0\\
		\textbf{H}_0
	\end{bmatrix},
\end{equation}
where
\begin{subequations}
	\begin{align} 
		\begin{cases}
			\label{aee_xyz} 
			(\rttensortwo{\alpha}_{ee})_{xx}  &= (\rttensortwo{\alpha}_{ee})_{yy} = (\rttensortwo{\alpha}_{ee})_{zz}=4\pi a^3 \epsilon_0  \epsilon \bigg[\frac{(\epsilon_{\rm m}-\epsilon)(2\mu+1) + 2\kappa^2 }{(\epsilon_{\rm m}+2\epsilon) (2\mu+1)-4\kappa^2 } \bigg]\\	
			(\rttensortwo{\alpha}_{ee})_{xy}  &= (\rttensortwo{\alpha}_{ee})_{yx}=(\rttensortwo{\alpha}_{ee})_{zx}=(\rttensortwo{\alpha}_{ee})_{xz}=(\rttensortwo{\alpha}_{ee})_{yz}=(\rttensortwo{\alpha}_{ee})_{zy}=0 
		\end{cases},
	\end{align}
\end{subequations}
\begin{subequations}
	\begin{align} 
		\begin{cases}
			\label{aem_xyz} 
			(\rttensortwo{\alpha}_{em})_{xx}  &= (\rttensortwo{\alpha}_{em})_{yy} = (\rttensortwo{\alpha}_{em})_{zz}= \frac{4\pi  a^3}{c}  \bigg[\frac{3  \epsilon \kappa}{(\epsilon_{\rm m}+2\epsilon) (2\mu+1)-4\kappa^2 } \bigg]\\	
			(\rttensortwo{\alpha}_{em})_{xy}  &= (\rttensortwo{\alpha}_{em})_{yx}=(\rttensortwo{\alpha}_{em})_{zx}=(\rttensortwo{\alpha}_{em})_{xz}=(\rttensortwo{\alpha}_{em})_{yz}=(\rttensortwo{\alpha}_{em})_{zy}=0 
		\end{cases},
	\end{align}
\end{subequations} 
\begin{subequations}
	\begin{align} 
		\begin{cases}
			\label{amm_xyz} 
			(\rttensortwo{\alpha}_{mm})_{xx}  &= (\rttensortwo{\alpha}_{mm})_{yy} = (\rttensortwo{\alpha}_{mm})_{zz}= -4\pi  a^3 \bigg[\frac{(\epsilon_{\rm m}+2\epsilon)(\mu-1)-2\kappa^2}{(\epsilon_{\rm m}+2\epsilon) (2\mu+1)-4\kappa^2 } \bigg]\\	
			(\rttensortwo{\alpha}_{mm})_{xy}  &= (\rttensortwo{\alpha}_{mm})_{yx}=(\rttensortwo{\alpha}_{mm})_{zx}=(\rttensortwo{\alpha}_{mm})_{xz}=(\rttensortwo{\alpha}_{mm})_{yz}=(\rttensortwo{\alpha}_{mm})_{zy}=0 
		\end{cases},
	\end{align}
\end{subequations} 
\begin{subequations}
	\begin{align} 
		\begin{cases}
			\label{amm_xyz} 
			(\rttensortwo{\alpha}_{me})_{xx}  &= (\rttensortwo{\alpha}_{me})_{yy} = (\rttensortwo{\alpha}_{me})_{zz}= 4\pi  a^3 c \epsilon_0 \bigg[\frac{3\kappa  \epsilon_{\rm m}}{(\epsilon_{\rm m}+2\epsilon) (2\mu+1)-4\kappa^2 } \bigg]\\	
			(\rttensortwo{\alpha}_{me})_{xy}  &= (\rttensortwo{\alpha}_{me})_{yx}=(\rttensortwo{\alpha}_{me})_{zx}=(\rttensortwo{\alpha}_{me})_{xz}=(\rttensortwo{\alpha}_{me})_{yz}=(\rttensortwo{\alpha}_{me})_{zy}=0 
		\end{cases}.
	\end{align}
\end{subequations}


\section{Effective optical constants of the metamaterial}
\begin{figure}[H]
	\centering
	\begin{center}
		\includegraphics[scale=0.3]{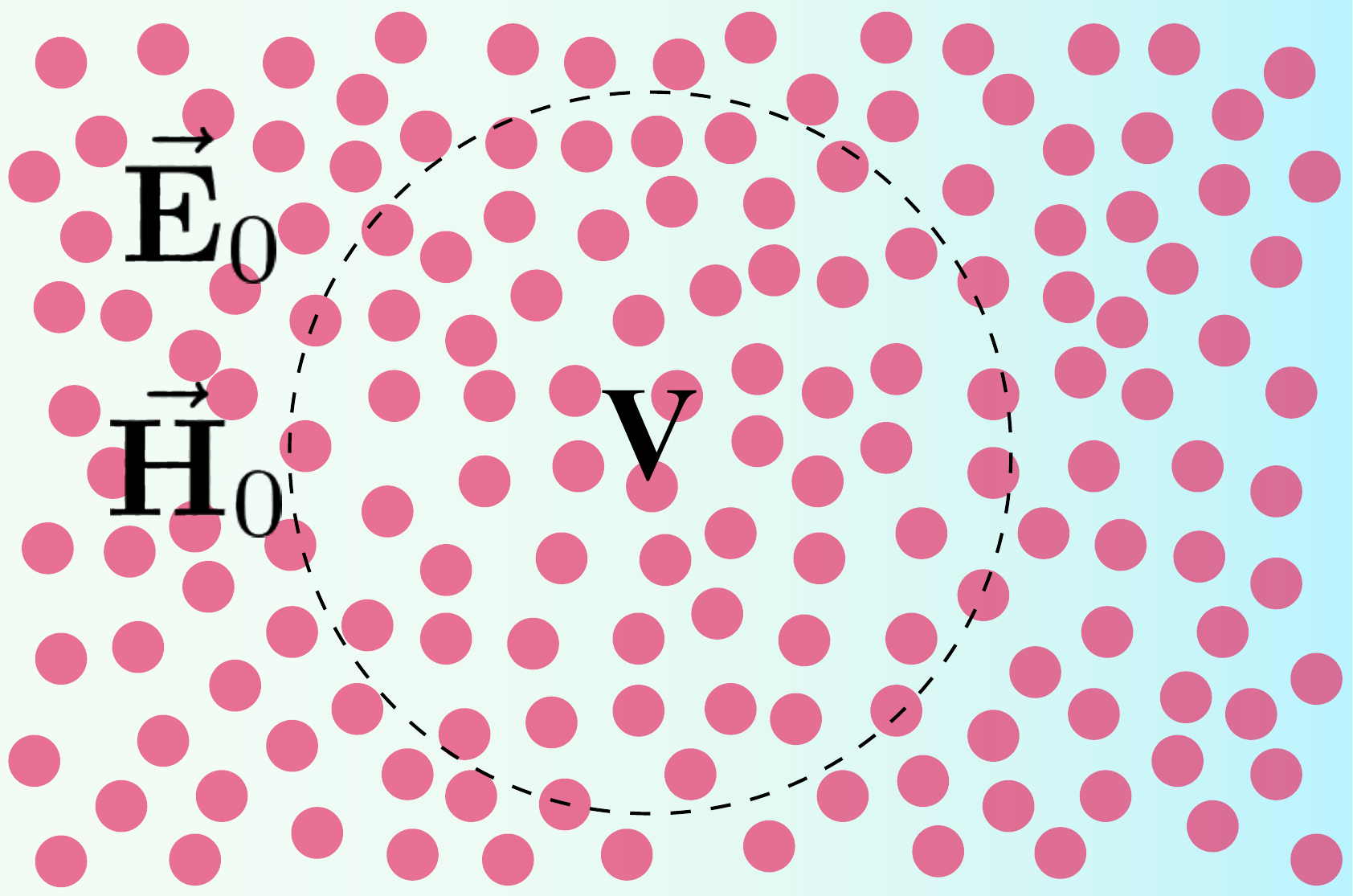}
		\captionsetup{labelformat=empty}
		\caption{FIG3. Schematic illustration of effective medium approach: $N$ metal spheres embedded in a chiral medium of volume $V$. $\textbf{E}_0$ and $\textbf{H}_0$ are the smoothed-out macroscopic fields neglecting the local field correction. The effective optical parameters can be evaluated from bulk polarisations.} 
		\label{fig3MG}
	\end{center}
\end{figure}
We consider a spherical volume $V$ containing $N$ number of NSs to calculate the effective optical constants. The total electric and magnetic polarisations ($\textbf{P}, \textbf{M}$) are given by the sum of induced dipole moments of the NS ($\textbf{P}_{\rm NS}, \textbf{M}_{\rm NS}$) and the reparixin polarisation ($\textbf{P}_{\rm CH}, \textbf{M}_{\rm CH}$), 
\begin{subequations}
	\begin{align}
		\begin{cases}
			\textbf{P}= \textbf{P}_{\rm NS}+\textbf{P}_{\rm CH}=  \frac{N}{V} \textbf{p} + \epsilon_0 (\epsilon-1) \textbf{E} - i\frac{\kappa}{c}\textbf{H}  \\	\\ 
			\textbf{M}= \textbf{M}_{\rm NS}+\textbf{M}_{\rm CH}=  \frac{N}{V} \textbf{m} +  (\mu-1) \textbf{H} + i \frac{\kappa}{c\mu_0}\textbf{E}
		\end{cases} 
	\end{align}
	\label{PMtot}
\end{subequations}
Now the induced dipole moments $\textbf{p}$ and $\textbf{m}$ can be expressed in terms of the fields $\textbf{E}_0$ and $\textbf{H}_0$ from Eq. \ref{pem1} as 
$\textbf{p}=\alpha_{ee}\textbf{E}_0 + i\alpha_{em}\textbf{H}_0$ and $\textbf{m}=\alpha_{mm}\textbf{H}_0 - i\alpha_{me}\textbf{E}_0$. Note that, adopting the Maxwell-Garnett homogenisation approach,  the macroscopic smoothed-out fields $\textbf{E}_0$ and $\textbf{H}_0$  are related to the local fields $\textbf{E}$ and $\textbf{H}$ as 
\begin{subequations}
	\begin{align}
		\begin{cases}
			\textbf{E}_0 &=  \textbf{E}+\frac{\textbf{P}}{3\epsilon_0}  \\	\\ 
			\textbf{H}_0 &= \textbf{H} - \frac{2\textbf{M}}{3}
			\label{EHin}
		\end{cases} .
	\end{align}
	\label{EHloc}
\end{subequations}
After simplifying Eqs. \ref{PMtot} and \ref{EHloc}, we obtain
\begin{equation} 
	\label{PMatrix} 
	\begin{bmatrix}
		\textbf{P}\\
		\textbf{M}
	\end{bmatrix}=	\frac{1}{\Gamma}\begin{bmatrix}
		1+\frac{2n}{3}\alpha_{mm} & & -i \frac{2n}{3}\alpha_{em}\\
		-i\frac{n\alpha_{me}}{3\epsilon_0} & & 1-\frac{n\alpha_{ee}}{3\epsilon_0}
	\end{bmatrix} \begin{bmatrix} n \alpha_{ee}+\epsilon_0 (\epsilon-1) & &i\left(n\alpha_{em}-\frac{\kappa}{c}\right) \\ i\left(-n\alpha_{me}+\frac{\kappa}{c\mu_0}\right) & & n\alpha_{mm} + (\mu-1)   
	\end{bmatrix} \begin{bmatrix}
		\textbf{E} \\ \textbf{H}
	\end{bmatrix},
\end{equation}
where $\Gamma = 1+(2n/3)\alpha_{mm}-(n/3\epsilon_0)\alpha_{ee}+(2n^2/9\epsilon_0)(\alpha_{em}\alpha_{me}-\alpha_{ee}\alpha_{mm})$ and $n=N/V$ is the number density of the NSs. Therefore, the constitutive relations for the bulk medium (NS+host chiral medium) can be calculated as
\begin{subequations}
	\begin{align}
		\begin{cases}
			\textbf{D} &=  \epsilon_0 \textbf{E} +  \textbf{P} \equiv \epsilon_0 \epsilon_{\rm eff} \textbf{E} -i \frac{\kappa_{\rm eff}}{c} \textbf{H}   \\	\\ 
			\textbf{B} &= \mu_0 (\textbf{H}+\textbf{M}) \equiv \mu_0 \mu_{\rm eff} \textbf{H} + i \frac{\kappa_{\rm eff}'}{c} \textbf{E}
			\label{EHin}
		\end{cases} 
	\end{align},
	\label{CR1}
\end{subequations}
leading to the expressions of the effective optical constants given in the main text.

\section{Calculation of ORD and ASF}
The wavenumbers $\beta_{\pm}$ for RCP ($s=+1$) and LCP ($s=-1$) excitation in the bulk chiral medium can be obtained from the curl equations of Maxwell equations
\begin{subequations}
	\begin{align}
		\label{kehout}
		\begin{cases}
			i\beta\hat{\textbf{k}} \times  \textbf{E}_{0} &=   ic \mu_0 \mu_{\rm eff} \textbf{H}_{0} - \kappa_{\rm eff} \textbf{E}_{0}\\	
			i\beta\hat{\textbf{k}} \times  \textbf{H}_{0} &=   -ic \epsilon_0 \epsilon_{\rm eff} \textbf{E}_{0} - \kappa_{\rm eff}' \textbf{H}_{0},
		\end{cases} 
	\end{align}
\end{subequations}
where $\boldsymbol{\beta}=\beta \hat{\textbf{k}}$ is the wavevector to be determined. Without loss of generalization, we take the propagation direction as $\hat{\textbf{k}}=\hat{\textbf{z}}$, and assuming circular polarization $\textbf{E}^{(\pm)}_{0}= (\hat{\textbf{x}}\pm i \hat{\textbf{y}})\rm E_0$, we have $\hat{\textbf{k}} \times \textbf{E}^{(\pm)}_{0}= \mp i\textbf{E}^{(\pm)}_{0}$. Therefore, from Eq. \ref{kehout} we obtain,
\begin{equation} 
	\beta_{\pm} = \frac{k_0}{2}\bigg[\mp (\kappa_{\rm eff} + \kappa_{\rm eff}')+ \sqrt{(\kappa_{\rm eff} - \kappa_{\rm eff}')^2 + 4\epsilon_{\rm eff} \mu_{\rm eff}   }\bigg],
	\label{betapm} 
\end{equation}
where $k_0=\omega/c$ is the free-space propagation constant.

\subsubsection{Calculating Optical rotatory dispersion (ORD)}
A linearly polarised incident wave $\textbf{E}_{\rm in} = \text{E}_0\hat{\textbf{e}}_x$ after passing through a distance of $z$ in the bulk chiral medium has the output field
\begin{equation} \label{propch} 
	\textbf{E}_{\rm out}(z) = \frac{\text{E}_0}{\sqrt{2}} \bigg[\hat{\textbf{e}}_x (e^{i\beta_+ z} + e^{i\beta_- z}) + i \hat{\textbf{e}}_y (e^{i\beta_+ z} - e^{i\beta_- z})  \bigg] .
\end{equation}
After simplifying, we obtain the ORD angle
\begin{equation} \label{ORD} 
	\theta = \frac{k_0 l}{2} \; \text{Re}[\kappa_{\rm eff} + \kappa_{\rm eff}'], 
\end{equation}
where $l$ is the length of the sample.

\subsubsection{Calculating Asymmetry Factor (ASF)}
The input and output fields are given by
\begin{eqnarray}
	\begin{cases}
		\bf{E_{\rm in}^{(\pm)}} & =  \frac{\hat{e}_x \pm i\hat{e}_y}{\sqrt{2}}\rm E_0 \\	\\
		\bf{E_{\rm out}^{(\pm)}} & =  \frac{\hat{e}_x \pm i\hat{e}_y}{\sqrt{2}}\rm E_0 e^{i\beta_{\pm} L}  
	\end{cases}.
\end{eqnarray}
In turn, the output intensity is given by
\begin{eqnarray}
	\rm {I_{\rm out}^{(\pm)}} & = |\rm E_0|^2 e^{-2\rm Im(\beta_{\pm}L)} ,
\end{eqnarray}
and the ASF factor is
\begin{equation}  
	\frac{\Delta I_{\rm ASF}}{\bar{I}} = 2 \frac{I_{\rm out}^+-I_{\rm out}^-}{I_{\rm out}^++I_{\rm out}^-}= 2\frac{e^{-2\rm Im(\beta_+)L}-e^{-2\rm Im(\beta_-)L}}{e^{-2\rm Im(\beta_+)L}+e^{-2\rm Im(\beta_-)L}} \approx -2 \text{tanh}\big[\text{Im}(\kappa_{\rm eff} + \kappa_{\rm eff}')k_0 L\big].
\end{equation}


\twocolumngrid

\end{document}